\documentclass[11pt,twoside,a4paper,cr,final]{cms-tdr}

\newcommand {\rootsNN}  {\ensuremath{\sqrt{s_{_{NN}}}}}
\newcommand {\PbPb}  {\mbox{PbPb}}
\newcommand {\AJ}       {\ensuremath{A_J}}
\newcommand {\pp}    {\mbox{pp}}
\newcommand {\dphi}     {\ensuremath{\Delta\phi}}

\cmsNoteHeader{HIN-11-004} % This is over-written in the CMS environment: useful as preprint no. for export versions

\begin{document}
\title{Jet measurements by the CMS experiment in pp and PbPb collisions}

\address[mit]{Massachusetts Institute of Technology, 77 Mass Ave, Cambridge, MA 02139-4307, USA\\}
\author[mit]{Christof Roland for the CMS collaboration}

\date{\today}

\abstract{
The energy loss of fast partons traversing the strongly interacting matter produced in high-energy nuclear collisions is one of the most interesting observables to probe the nature of the produced medium. The multipurpose Compact Muon Solenoid (CMS) detector is well designed to measure these hard scattering processes with its high resolution calorimeters and high precision silicon tracker. Analyzing data from pp and PbPb collisions at a center-of-mass energy of 2.76 TeV parton energy loss is observed as a significant imbalance of dijet transverse momentum. To gain further understanding of the parton energy loss mechanism the redistribution of the quenched jet energy was studied using the transverse momentum balance of charged tracks projected onto the direction of the leading jet. In contrast to pp collisions, a large fraction the momentum balance for asymmetric jets is found to be carried by low momentum particles at large angular distance to the jet axis. Further, the fragmentation functions for leading and subleading jets were reconstructed and were found to be unmodified compared to measurements in pp collisions. The results yield a detailed picture of parton propagation in the hot QCD medium.
}

\hypersetup{%
pdfauthor={CMS Collaboration},%
pdftitle={Charmonium production measured in PbPb and pp collisions by CMS},%
pdfsubject={CMS},%
pdfkeywords={CMS, physics, heavy-ions, jet quenching}}

\maketitle %maketitle comes after all the front information has been supplied

\section{Introduction}

\begin{figure}[t!]
\begin{center}
\includegraphics[width=1.0\textwidth]{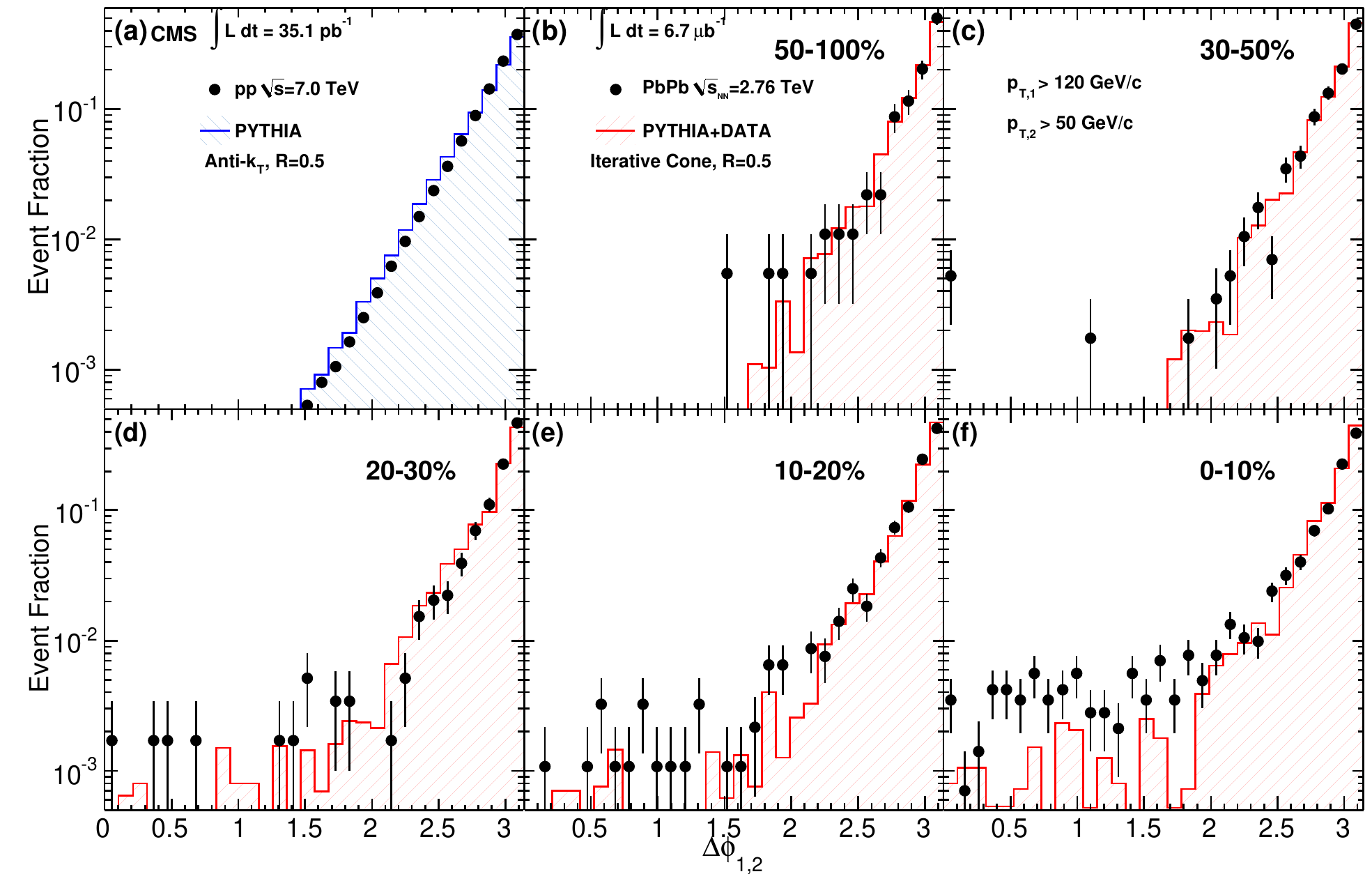}
\end{center}
\caption{$\Delta \phi_{12}$ distributions for leading jets of $p_{\mathrm{T},1} > 120$~\GeVc\ 
with subleading jets of $p_{\mathrm{T},2} >  50$~\GeVc for 
7 TeV pp collisions (a) and 2.76 TeV PbPb collisions in several centrality bins:
(b) 50--100\%, (c) 30--50\%, (d) 20--30\%, (e) 10--20\% and (f) 0--10\%.
Data are shown as black points, while the histograms show (a) {\sc{pythia}} events  and  (b)-(f) {\sc{pythia}} events embedded into PbPb data.  
The error bars show the statistical uncertainties.}
\label{fig:dphi50}
\end{figure}

Heavy ion collisions at the Large Hadron Collider (LHC) are expected to produce matter at 
energy densities exceeding any previously explored in experiments conducted at particle 
accelerators. One of the key experimental signatures suggested to study the quark-gluon plasma(QGP), which is expected to be formed at these high energy densities, is the energy loss
 of high-transverse-momentum partons passing through the medium~\cite{Bjorken:1982tu}.
This parton energy loss is often referred to as ``jet quenching''.
The energy lost by a parton provides fundamental information on the thermodynamical
and transport properties of the traversed medium.

\section{Data set and experimental method}
\begin{figure}[t!]
\begin{center}
\includegraphics[width=1.0\textwidth]{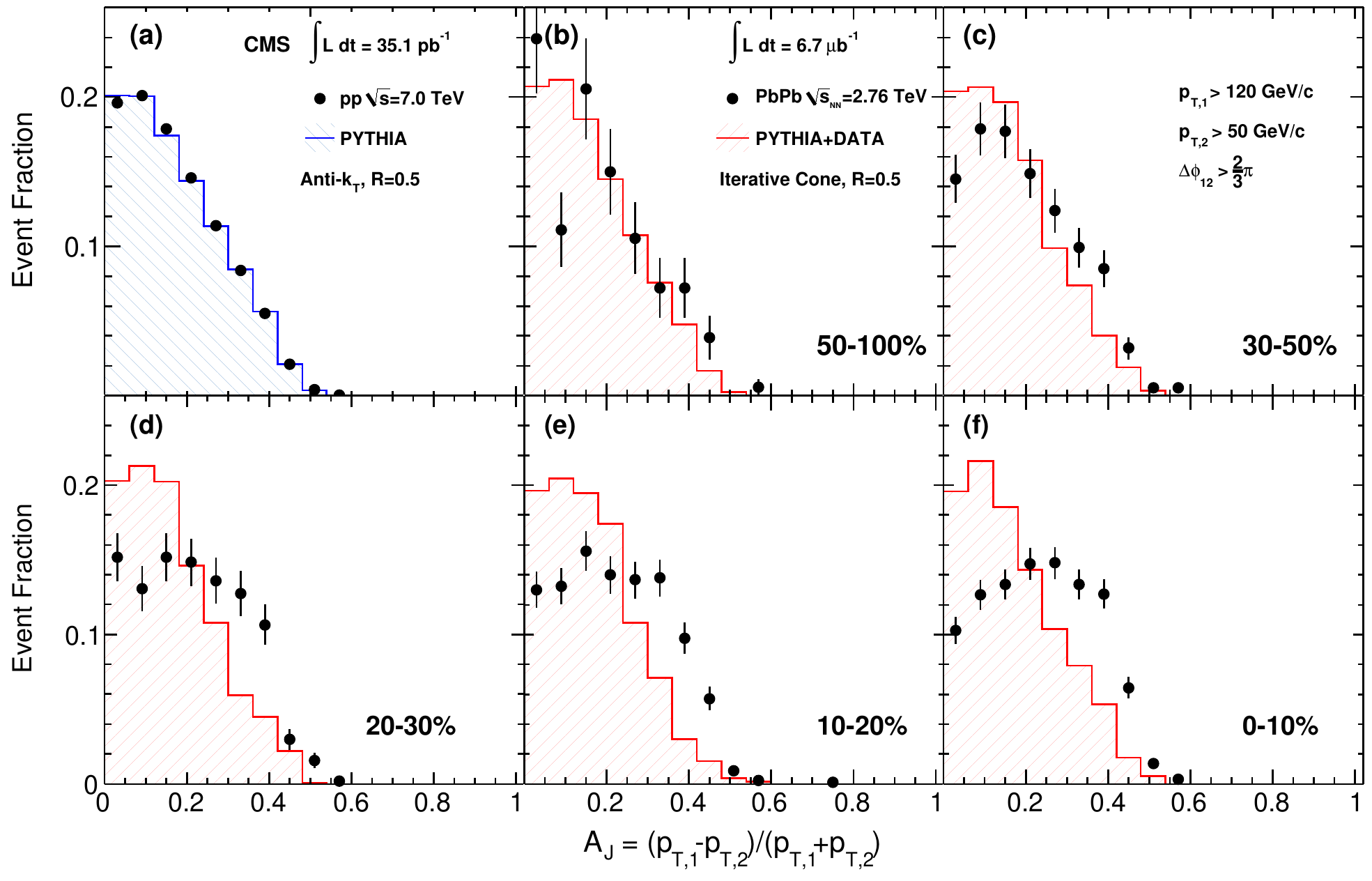}
\caption{Dijet asymmetry ratio, $A_{J}$, for leading jets of $p_{\mathrm{T},1}> $ 120~\GeVc, subleading jets of $p_{\mathrm{T},2}> $50~\GeVc\, and $\Delta\phi_{12}>2\pi/3$  for 7 TeV pp collisions (a) and 2.76 TeV \PbPb\ collisions in several centrality bins:  (b) 50--100\%, (c) 30--50\%, (d) 20--30\%, (e) 10--20\% and (f) 0--10\%.
Data are shown as black points, while the histograms show  (a) {\sc{pythia}} events and  (b)-(f) {\sc{pythia}} events embedded into \PbPb\ data.  
The error bars show the statistical uncertainities.}
\label{fig:JetAsymm}                  
\end{center}
\end{figure}
The dijet analysis presented in this paper was performed using the data collected from pp and 
PbPb collisions at a nucleon-nucleon center-of-mass energy of \rootsNN$=2.76$~TeV 
at the Compact Muon Solenoid (CMS) detector.  The CMS detector has 
a solid angle acceptance of nearly 4$\pi$ and is designed to measure jets and energy flow, an ideal feature for studying heavy ion collisions. The CMS detector is described in detail
elsewhere~\cite{bib_CMS}.
This analysis is based on a total integrated PbPb (pp) luminosity of 6.7~$\mu$b$^{-1}$ (225~nb$^{-1}$). 
The jets used for analysis were reconstructed with two dif\-ferent algorithms.
The results related to dijet correlations uses jets reconstructed from calorimeter 
energies using an iterative cone algorithm with a cone radius of 0.5. 
The fragmentation function results use jets reconstructed using the CMS particle flow 
algorithm, which combines information from all detector systems \cite{MattTalk:2011}
and the anti-k$_T$ algorithm, as encoded in the FastJet 
framework, with a resolution parameter of R = 0.3 \cite{Cacciari:2008gp}.
For both algorithms the underlying event is subtracted using the method 
described in \cite{Kodolova:2007hd} and the reconstructed jet energies are corrected 
for the detector response based on pythia \cite{bib_pythia} simulations.
Charged particles are reconstructed in the CMS Silicon tracking system \cite{bib_CMS}.

\section{Dijet Correlations}
 For this analysis dijets are selected with a leading jet of $p_{\mathrm{T},1}> $ 120~\GeVc, a subleading jet of $p_{\mathrm{T},2}> $50~\GeVc\, both within $|\eta| <$ 2 and with an opening angle of $\Delta\phi_{12}>2\pi/3$, unless otherwise noted. More details on the analysis presented in this section can be found in \cite{Chatrchyan:2011sx}.

\subsection{Dijet azimuthal correlations}
\begin{figure}[th!]
  \begin{center}
    \includegraphics[width=1.0\textwidth]{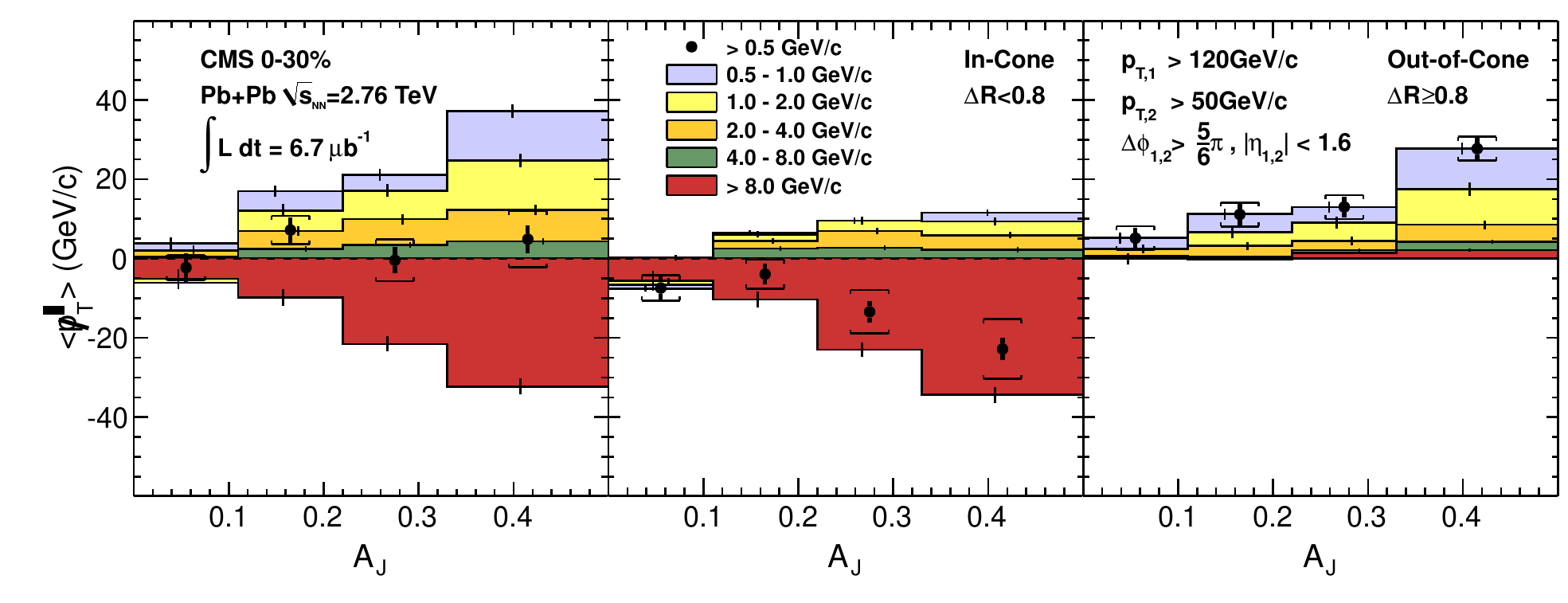}
    \caption{Left Panel: Average missing transverse momentum, $\langle \displaystyle{\not} p_{\mathrm{T}}^{\parallel} \rangle$, for tracks with \pt $> 0.5$~\GeVc, projected onto the leading jet axis (solid circles). The $\langle \displaystyle{\not} p_{\mathrm{T}}^{\parallel} \rangle$ values are shown as a function of dijet asymmetry
$A_J$ for 0--30\% central events. 
Middle Panel: The $\langle \displaystyle{\not} p_{\mathrm{T}}^{\parallel} \rangle$ values as a function of $A_J$ inside ($\Delta R < 0.8$) one of the leading or subleading jet cones. Right panel: $\langle \displaystyle{\not} p_{\mathrm{T}}^{\parallel} \rangle$ outside ($\Delta R > 0.8$) the leading and subleading jet cones.
For the solid circles, vertical bars and brackets represent the statistical and systematic 
uncertainties, respectively. Colored bands show the contribution to $\langle \displaystyle{\not} p_{\mathrm{T}}^{\parallel} \rangle$ for five ranges of track \pt.
For the individual \pt ranges, the statistical uncertainties are shown as vertical bars.
}
    \label{fig:MissingpT}
  \end{center}
\end{figure}
One possible medium ef\-fect on the dijet properties is a change of the back-to-back alignment of 
the two partons. This can be studied using the event-normalized dif\-ferential dijet distribution, ($1/N$)($dN / d\Delta \phi_{12}$), 
%($1/\text{N)(}\text{dN} / \text{d} \Delta \phi_{12}$), 
versus $\Delta \phi_{12}$. 
Figure~\ref{fig:dphi50} shows distributions of $\Delta \phi_{12}$ between leading and subleading jets which pass the respective \pt\ selections.
 %for leading jets of $p_{\mathrm{T},1} > 120$~\GeVc\ and subleading jets of $p_{\mathrm{T},2} >  50$~\GeVc. 
In Figure~\ref{fig:dphi50} (a), the dijet $\Delta \phi_{12}$ distributions are plotted for 7~TeV \pp\ data in comparison to 
the corresponding {\sc{pythia}} simulations using the anti-$k_{\rm T}$ algorithm for jets based on calorimeter information. 
{\sc{pythia}} provides a good description of the experimental data, with slightly larger tails seen in the {\sc{pythia}} simulations. 
Figures~\ref{fig:dphi50}~(b)-(f) show the dijet $\Delta \phi_{12}$ distributions for 
\PbPb\ data in five centrality bins, compared to {\sc{pythia+data}} simulations. 
For all centrality bins a good agreement with the {\sc{pythia+data}} reference is found for the bulk of the data, especially for \dphi$_{12} > 2$. 
%
%The three centrality bins spanning 0--30\% show an excess of events with azimuthally 
%misaligned dijets ($\dphi_{12} \lesssim 2$), compared with more
%peripheral events. A similar trend is seen for the {\sc{pythia+data}} simulations, 
%although the fraction of events with  azimuthally misaligned dijets is smaller in the 
%simulation. The centrality dependence of the azimuthal correlation in {\sc{pythia+data}} can
%be understood as the result of the increasing fake-jet rate and the
%drop in jet reconstruction efficiency near the 50~\GeVc\ threshold from 95\% for
%peripheral events to 88\% for the most central events. In \PbPb\ data, 
%this effect is magnified since low-\pt away-side jets can undergo a sufficiently 
%large energy loss to fall below the 50~\GeVc\ selection criteria.
\subsection{Dijet momentum balance}
\label{sec:asymmetry}
To characterize the dijet momentum balance (or imbalance) quantitatively, we use the 
asymmetry ratio,
\begin{equation}
\label{eq:aj} 
A_J = \frac{p_{\mathrm{T},1}-p_{\mathrm{T},2}}{p_{\mathrm{T},1}+p_{\mathrm{T},2}}~,
\end{equation} 
where the subscript $1$ always refers to the leading jet, so that $A_J$ is positive by 
construction. 
In Fig.~\ref{fig:JetAsymm} (a),  the $A_J$ dijet asymmetry observable calculated 
by {\sc{pythia}} is compared to \pp\  data at $\sqrt{s}$ = 7~TeV. 
Data and event generator are found to be in excellent agreement.
%This observation, as well as the good agreement between {\sc{pythia+data}} and the most peripheral \PbPb\ data shown in Fig.~\ref{fig:JetAsymm} (b), suggests that {\sc{pythia}} at $\sqrt{s}$ = 2.76~TeV can serve as a good reference for the dijet imbalance analysis in \PbPb\ collisions. 
%
The centrality dependence of $A_J$ for \PbPb\ collisions can be seen in
Figs.~\ref{fig:JetAsymm}~(b)-(f), in comparison to {\sc{pythia+data}} simulations.
The dijet momentum balance exhibits a dramatic change in shape for the most central 
collisions. In contrast, the {\sc{pythia}} simulations only
exhibit a modest broadening, even when embedded in the highest multiplicity \PbPb\ events. 
%Central \PbPb\ events show a significant deficit of events in which the momenta of leading and subleading jets are balanced and a significant excess of unbalanced pairs.
This observation is consistent with a degradation of the parton energy, or jet quenching, in the
medium produced in central \PbPb\ collisions.

\subsection{Overall momentum balance of dijet events}
Information about the overall momentum balance in the dijet events can be obtained using 
the projection of missing \pt\ of reconstructed charged tracks onto the leading jet axis. 
For each event, this projection was calculated as 

\begin{equation}
\displaystyle{\not} p_{\mathrm{T}}^{\parallel} = 
\sum_{\rm i}{ -p_{\mathrm{T}}^{\rm i}\cos{(\phi_{\rm i}-\phi_{\rm Leading\ Jet})}},
\end{equation}

where the sum is over all tracks with \pt $> 0.5$~\GeVc\ and $|\eta| < 2.4$. 
The results were then averaged over events to obtain $\langle \displaystyle{\not} 
p_{\mathrm{T}}^{\parallel} \rangle$.
The leading and subleading jets were required to have $|\eta| < 1.6$.
The left panel of Fig.~\ref{fig:MissingpT} shows $\langle \displaystyle{\not} p_{\mathrm{T}}^{\parallel} \rangle$ as a function of \AJ\ for the 0--30\% most central \PbPb\ collisions. 
Using tracks with $|\eta| < 2.4$ and \pt $> 0.5$~\GeVc, one sees
that, even for events with large observed dijet asymmetry, the momentum balance of the events, 
shown as solid circles, is recovered within uncertainties.
%
%This shows that the dijet momentum imbalance is not related to undetected activity in the event due to instrumental (e.g. gaps or inefficiencies in the calorimeter) or physics (e.g. neutrino production) effects. 
%
The figure also shows the contributions to 
$\langle \displaystyle{\not} p_{\mathrm{T}}^{\parallel} \rangle$ for five transverse 
momentum ranges from 0.5--1~\GeVc\ to \pt $> 8$~\GeVc. The vertical bars for each range 
denote statistical uncertainties. 
A large negative contribution to $\langle \displaystyle{\not} p_{\mathrm{T}}^{\parallel} \rangle$ (i.e., in the direction of the leading jet) by the \pt $> 8$~\GeVc\ range is balanced by the combined contributions of low-momentum tracks from the 0.5--2~\GeVc\ regions.  

%In peripheral \PbPb\ data, the contribution of 0.5--2~\GeVc\ tracks relative to that from 4--8~\GeVc tracks is somewhat enhanced compared to the simulation.
%In central \PbPb\ events, the relative contribution of low and intermediate-\pt\  tracks is actually the opposite of that seen in {\sc{pythia+hydjet}}. 
%In data, the 4--8~\GeVc\ region makes almost no contribution to the overall momentum balance, while a large fraction of the negative imbalance from high \pt\ is recovered in low-momentum tracks.

Further insight into the radial dependence of the momentum balance can be gained by studying 
$\langle \displaystyle{\not} p_{\mathrm{T}}^{\parallel} \rangle$ separately for tracks inside 
cones of size $\Delta R = 0.8$ around the leading and subleading jet axes, and for tracks 
outside of these cones. The results of this study are shown in Fig.~\ref{fig:MissingpT} for the in-cone balance (middle panel) and out-of-cone balance (right panel).
As the underlying \PbPb\ event is not $\phi$-symmetric on an event-by-event basis, the back-to-back requirement was tightened to $\dphi_{12} > 5 \pi/6$ for this study.

One observes that the in-cone imbalance of $\langle \displaystyle{\not} p_{\mathrm{T}}^{\parallel} \rangle \approx
-20$~\GeVc\ is found for the $\AJ > 0.33$ selection, which is balanced by a corresponding 
out-of-cone imbalance of  $\langle \displaystyle{\not} p_{\mathrm{T}}^{\parallel} \rangle \approx 20$~\GeVc. In the \PbPb\ data the out-of-cone contribution is carried almost entirely  by tracks with $0.5 < $\pt$ < 4$~\GeVc\ .

\section{Fragmentation Functions}
\begin{figure}[th!]
\begin{center}
\includegraphics[width=0.9\textwidth]{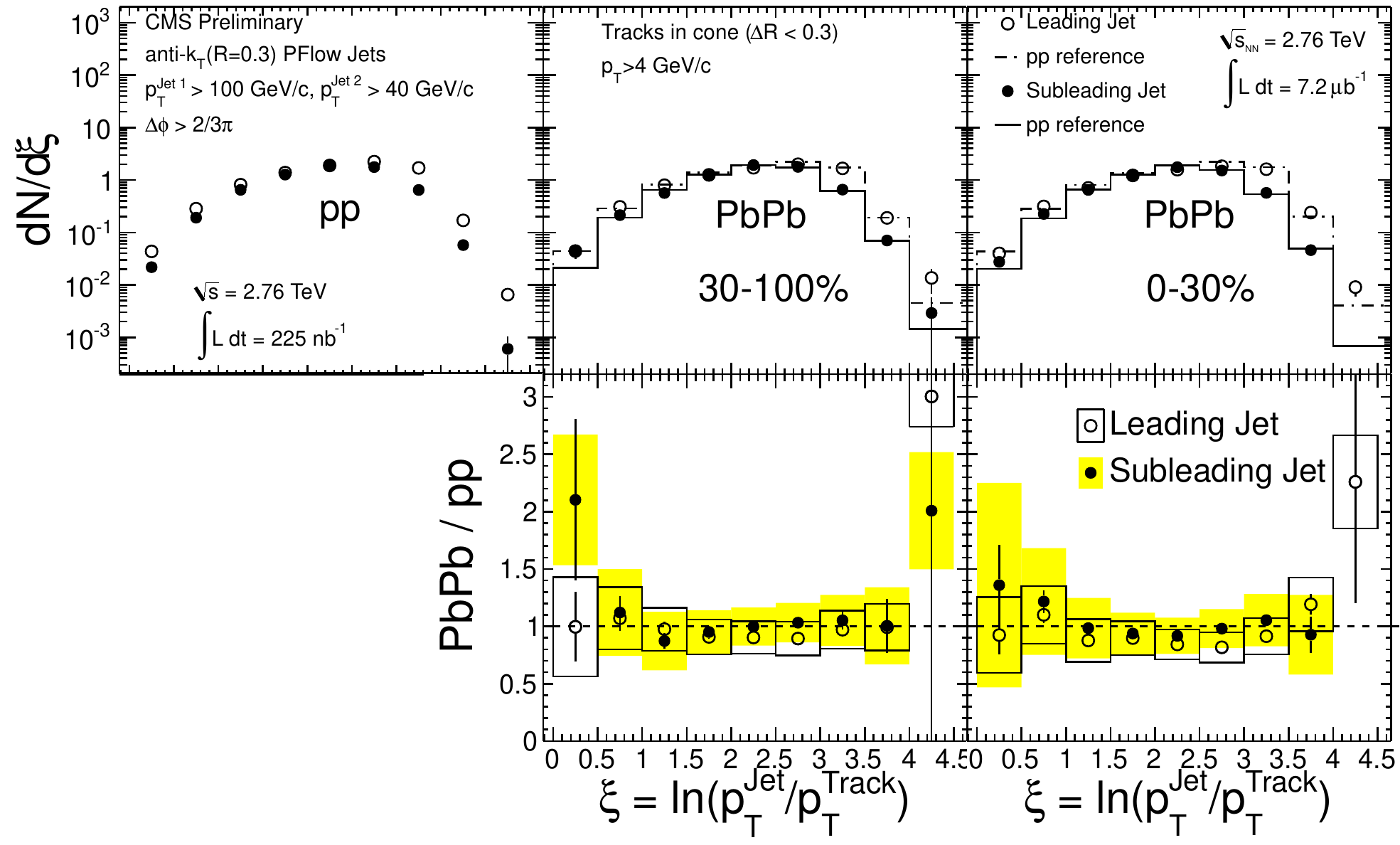}
\includegraphics[width=0.9\textwidth]{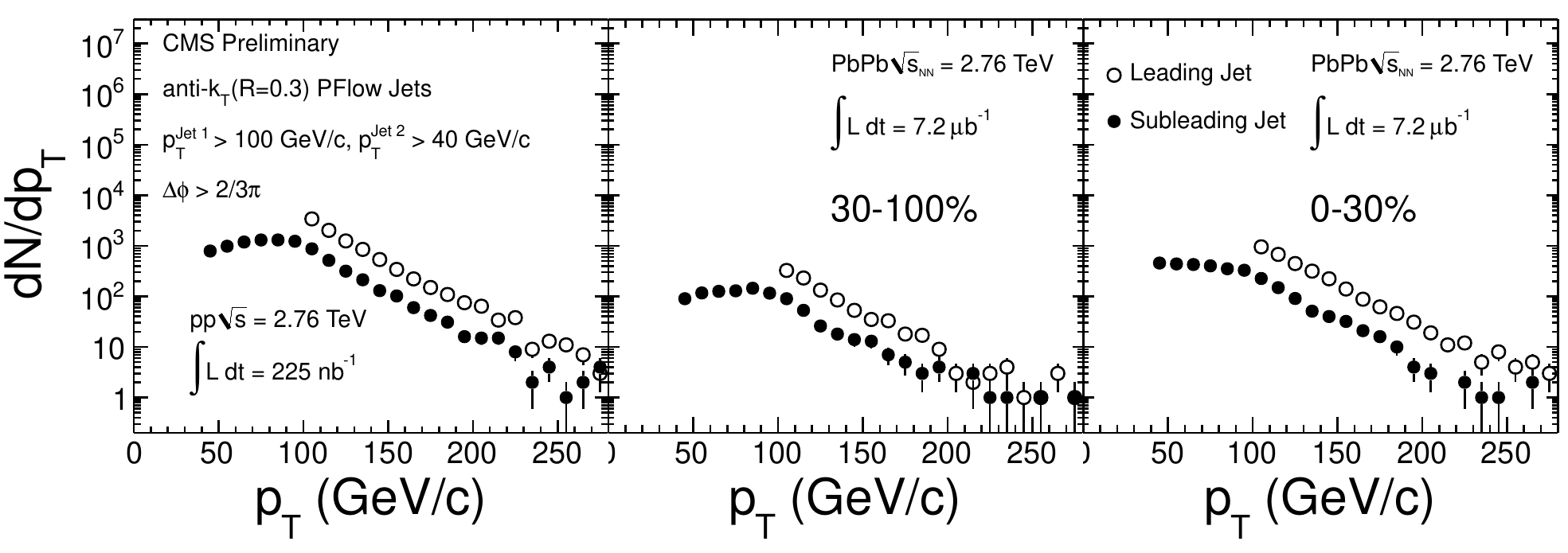}
\end{center}
\caption{Fragmentation functions reconstructed in pp, peripheral PbPb, and central PbPb data, with the leading (open circles) and subleading (solid points) jets shown in the top row. The middle row is the ratio of each PbPb fragmentation function to its pp reference. The pp reference is constructed by convoluting the pp fragmentation with jet \pt\ resolution ef\-fects of PbPb collisions and reweighting with jet \pt. Error bars are statistical, the hollow boxes represent the systematic uncertainty for the leading jet, and yellow boxes are the systematic uncertainty for the subleading jet. The bottom row shows the jet \pt distributions of the samples (pp, peripheral PbPb, and central PbPb). It is important to note that jet \pt\ distributions of leading and subleading jets are dif\-ferent.}
\label{fig:ffLeadSublead_pp_ppDiv}
\end{figure}
\begin{figure}[h]
\begin{center}
\includegraphics[width=0.9\textwidth]{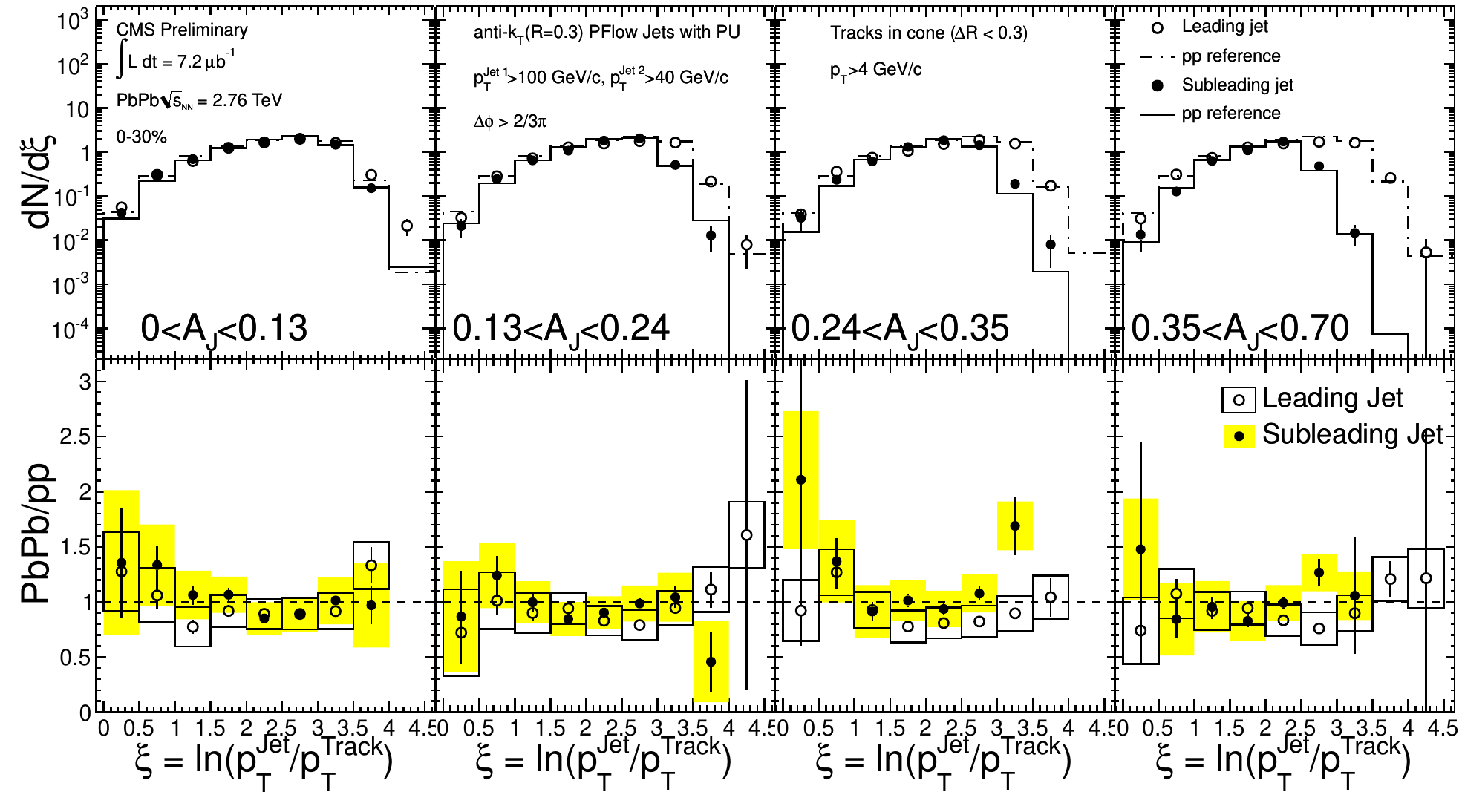}
\includegraphics[width=0.9\textwidth]{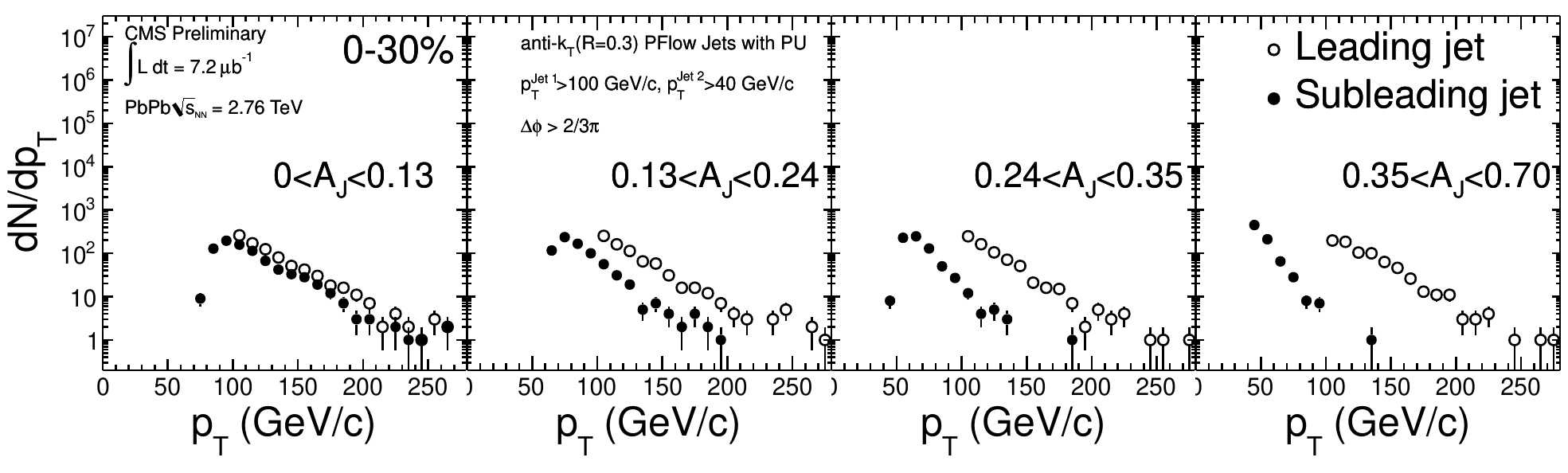}
\end{center}
\caption{Fragmentation functions in $\AJ$ bins, reconstructed in central PbPb and pp reference for the leading 
(open circles) and subleading (solid points) jets. The middle row shows the ratio of each fragmentation function to 
its convoluted pp reference. The systematic uncertainty is represented by an hollow box (leading jet) or yellow box 
(subleading jet). Error bars shown are statistical. The bottom row shows the jet \pt\ distributions. 
The reference distributions are reweighted to match the jet \pt\ distributions in PbPb.
}
\label{fig:ffLeadSublead_centralAJ_ppDiv}
\end{figure}
Fragmentation functions are measured by correlating tracks within the jet cones of reconstructed dijets with the jet axis for each jet. For this analysis dijets are selected with a
leading jet of $p_{\mathrm{T},1}> $ 100~\GeVc, a subleading jet of $p_{\mathrm{T},2}> $40~\GeVc\, within $|\eta| <$ 2 and an opening angle of $\Delta\phi_{12}>2\pi/3$. The reconstructed jets are corrected to the generator final state particle level and the reconstructed charged particles are corrected for tracking ef\-ficiency. The fragmentation function is a function of $\xi$, defined as:
\begin{equation}
\label{eq:xi} 
\xi = - \ln \mathrm{z} = - \ln \frac{p_{T}^{track}}{p_{T}^{jet}},
\end{equation}
where \pt$^{jet}$ is the transverse momentum of the reconstructed jet, and \pt$^{track}$ is the transverse momentum of tracks within the jet cone ($\Delta R=\sqrt{\Delta\phi^2+\Delta \eta^2} = 0.3$ around the jet axis). Tracks are selected with \pt$^{track} > 4 \GeVc$, which restricts this measurement to the low $\xi$ region of the fragmentation function. For the \PbPb\ analysis, this track selection minimizes the contribution of the underlying event to the jet fragmentation function.
%
%To establish a reference for the measurement in \PbPb\ collisions we reconstruct fragmentation functions in pp collisions at the same center of mass energy. 
%
Figure~\ref{fig:ffLeadSublead_pp_ppDiv} shows the reconstructed leading and subleading jet fragmentation functions (top row) in a cone of $\Delta R = 0.3$ around the respective jet axis for \pp\ collisions (left panel) peripheral \PbPb\ (middle panel) and central \PbPb\ (right panel). The corresponding jet \pt\ distributions are shown in the bottom row. 

For a direct comparison between pp and PbPb, the pp data has to include the resolution deterioration due to the underlying event fluctuations seen in PbPb. For this purpose the pp jet \pt\ has been artificially smeared by the fluctuations observed in PbPb collisions taking into account the mean and RMS of the fluctuations as well as the correlation of the background level seen in back-to-back jet cones and the centrality dependence of the fluctuation parameters. 
The ratio between the PbPb and fluctuation convoluted pp reference fragmentation functions 
are shown in the middle rows for peripheral and central PbPb collisions.
In both cases, the fragmentation functions of the leading and subleading jet are in agreement with pp collisions. 
%This agreement of the hard fragmentation pattern of partons that have traversed dif\-ferent path lengths in the nuclear medium is an indication that the fragmentation properties of partons are not modified by the energy loss process. 

\subsection{Fragmentation Functions in Bins of $\AJ$}
To study a potential ef\-fect of parton energy loss on the fragmentation properties of partons in more detail, we divide the data sample into classes of dijet imbalance. Four $\AJ$ bins are chosen, which split the data sample into approximately equal parts: 0 $< \AJ <$ 0.13, 0.13 $< \AJ <$ 0.24, 0.24 $< \AJ <$ 0.35, and 0.35 $< \AJ <$ 0.70. The fragmentation functions are reconstructed separately for leading and subleading jets in central events.
Figure \ref{fig:ffLeadSublead_centralAJ_ppDiv} shows the fragmentation functions in bins of increasing dijet imbalance (\AJ) from left to right (top row), the corresponding ratio to the fluctuation convoluted pp reference (middle row) and the jet \pt\ distribution for leading and subleading jet (bottom row).
For both leading and subleading jets, independent of the dijet imbalance bin, the \PbPb\ fragmentation functions closely resemble those of the \pp\ reference. 

\section{Summary}
\label{sec:summary}

The CMS detector has been used to study jet production in PbPb collisions at 
\rootsNN$=\,2.76~TeV$. No visible modification of the dijet azimuthal 
correlation due to the nuclear medium was observed while a strong increase 
in the fraction of highly unbalanced jets has been seen in central PbPb 
collisions compared with peripheral collisions and model calculations. 
This observation is consistent with a high degree of jet quenching in the 
produced matter. A large fraction of the momentum balance of the 
unbalanced jets is carried by low-\pt\ particles at large radial 
distance to the jet axis.
The hard component of fragmentation functions reconstructed 
in \PbPb\ collisions for different event centrality and dijet 
imbalance $\AJ$ exhibit a universal behavior closely resembling 
jets of the same reconstructed energy fragmenting without having traversed 
a nuclear medium, as seen in the comparison with pp collisions.
These results provide qualitative constraints on the nature of the jet 
modification in \PbPb\ collisions and quantitative input to models of 
the transport properties of the medium created in these collisions.


\providecommand{\href}[2]{#2}\begingroup\raggedright\begin{thebibliography}{1}%
\makeatletter
\providecommand{\hrefCMSnoop }[0]{\@secondoftwo}%
\makeatother

\bibitem{Bjorken:1982tu}
\hrefCMSnoop {} {J.~D. Bjorken, ``Energy loss of energetic partons in
  QGP:possible extinction of high $p_T$ jets in hadron-hadron collisions'',}.
  FERMILAB-PUB-82-059-THY.

\bibitem{bib_CMS}
\hrefCMSnoop {} {{ CMS} Collaboration, ``The CMS experiment at the CERN LHC'',}
  \textit{ JINST} \textbf{ 3} (2008) S08004.
  \href{http://dx.doi.org/10.1088/1748-0221/3/08/S08004}{\texttt{
  doi:10.1088/1748-0221/3/08/S08004}}.

\bibitem{MattTalk:2011}
\hrefCMSnoop {} {M.~Nguyen, ``{Jet reconstruction with particle flow in
  heavy-ion collisions with CMS}'',} \textit{ These proceedings} (2011).

\bibitem{Cacciari:2008gp}
\hrefCMSnoop {} {M.~Cacciari, G.~P. Salam, and G.~Soyez, ``{The anti-$k_t$ jet
  clustering algorithm}'',} \textit{ JHEP} \textbf{ 04} (2008) 063,
  \href{http://www.arXiv.org/abs/0802.1189}{\texttt{ arXiv:0802.1189}}.
\href{http://dx.doi.org/10.1088/1126-6708/2008/04/063}{\texttt{
  doi:10.1088/1126-6708/2008/04/063}}.
%%CITATION = 0802.1189;%%.

\bibitem{Kodolova:2007hd}
\hrefCMSnoop {} {O.~Kodolova, I.~Vardanian, A.~Nikitenko{ et~al.}, ``{The
  performance of the jet identification and reconstruction in heavy ions
  collisions with CMS detector}'',} \textit{ Eur. Phys. J.} \textbf{ C50}
  (2007) 117.
\href{http://dx.doi.org/10.1140/epjc/s10052-007-0223-9}{\texttt{
  doi:10.1140/epjc/s10052-007-0223-9}}.
%%CITATION = EPHJA,C50,117;%%.

\bibitem{bib_pythia}
\hrefCMSnoop {} {T.~Sj{\"o}strand, S.~Mrenna, and P.~Skands, ``{PYTHIA 6.4
  Physics and Manual}'',} \textit{ JHEP} \textbf{ 05} (2006) 026 (tune D6T with
  PDFs CTEQ6L1 used), \href{http://www.arXiv.org/abs/hep-ph/0603175}{\texttt{
  arXiv:hep-ph/0603175}}.

\bibitem{Chatrchyan:2011sx}
\hrefCMSnoop {} {{ CMS} Collaboration, ``{Observation and studies of jet
  quenching in PbPb collisions at nucleon-nucleon center-of-mass energy = 2.76
  TeV}'',} \href{http://www.arXiv.org/abs/1102.1957}{\texttt{
  arXiv:1102.1957}}.

\end{thebibliography}\endgroup
\end{document}